\documentclass[a4paper,11pt]{article}
\usepackage{jheppub}
\usepackage[T1]{fontenc}
\usepackage{amsmath}
\usepackage{amssymb}
\usepackage{graphicx}
\usepackage{esint}
\usepackage{epsfig}
\usepackage{amsfonts}
\usepackage{adjustbox,lipsum}

\def\beq{\begin{eqnarray}}
\def\eeq{\end{eqnarray}}
\def\bea{\begin{eqnarray}}
\def\eea{\end{eqnarray}}
\definecolor{blue}{cmyk}{1,0.9,0,0.3}

\title{\color{blue}A Radiative Model for the Weak Scale and Neutrino Mass via Dark Matter}
\author[a,b]{Amine Ahriche}\author[c]{Kristian~L.
McDonald}\author[d]{Salah Nasri} \affiliation[a]{Department of
Physics, University of Jijel, PB 98 Ouled Aissa, DZ-18000 Jijel,
Algeria.} \affiliation[b]{The Abdus Salam International Centre for
Theoretical Physics, Strada Costiera 11, I-34014, Trieste, Italy.}
\affiliation[c]{ARC Centre of Excellence for Particle Physics at the Terascale,\\
School of Physics, The University of Sydney, NSW 2006, Australia.}
\affiliation[d]{Department of Physics, UAE University, P.O. Box
17551, Al-Ain, United Arab Emirates.} \emailAdd{aahriche@ictp.it,
kristian.mcdonald@sydney.edu.au, snasri@uaeu.ae.ac}

\abstract{We present a three-loop model of neutrino mass in which
both the weak scale and neutrino mass arise as radiative effects. In
this approach, the scales for electroweak symmetry breaking, dark
matter, and the exotics responsible for neutrino mass, are related
due to an underlying scale-invariance. This motivates the
otherwise-independent $\mathcal{O}(\mathrm{TeV})$ exotic masses
usually found in three-loop models of neutrino mass. We demonstrate
the existence of viable parameter space and show that the model can
be probed at colliders, precision experiments, and dark matter
direct-detection experiments.}

\begin{document}

\maketitle

\flushbottom

\section{Introduction\label{sec:introduction}}

Radiative symmetry breaking~\cite{Coleman:1973jx} offers an interesting
alternative to the conventional Higgs mechanism. In this approach,
calculable weakly-coupled radiative effects induce symmetry breaking in
classically scale-invariant theories, thereby giving birth to mass --- a
process known as dimensional transmutation. When applied to the Standard
Model (SM), it is well known that radiative symmetry breaking is not viable,
due to the destabilizing influence of the heavy top quark. However, the SM
is known to be incomplete, due to e.g.~an absence of massive neutrinos and
the need to incorporate dark matter (DM). It is therefore interesting to
consider the viability of radiative symmetry breaking within SM extensions.

The addition of massive neutrinos and DM to the SM likely requires new
degrees of freedom. When considering radiative symmetry breaking, there are
a number of relevant considerations that can guide the choice of beyond-SM
fields. The destabilizing radiative corrections from the top quark can be
overcome by bosonic degrees of freedom with mass $\gtrsim 200$~\textrm{GeV}.
In principle these states could be much heavier than the \textrm{TeV} scale.
However, radiative symmetry breaking typically introduces a single scale
into a theory, with other mass and symmetry breaking scales related to this
scale.\footnote{%
The exceptions being when a theory also contains a confining gauge sector,
as with QCD in the SM, or a completely decoupled hidden sector possessing
its own symmetry breaking and/or confining pattern.} Consequently both the
electroweak scale and the mass scale for exotics may be related via
dimensionless parameters. Thus, absent hierarchically small parameters~\cite%
{Kobakhidze:2014afa}, one anticipates exotics with $\mathcal{O}(\mathrm{TeV}%
) $ masses.

In the LHC era, \textrm{TeV} scale exotics are of particular interest.
However, efforts to generate tiny neutrino masses via weak-scale exotics can
struggle to achieve the necessary mass-suppression, relative to the weak
scale, without invoking tiny couplings. Perhaps the most obvious exception
are models with radiative neutrino mass, as the inherent loop-suppression in
such models can motivate lighter new physics. From this perspective,
three-loop models of neutrino mass are particularly compelling, as the new
physics is expected to be $\mathcal{O}(\mathrm{TeV})$.

These considerations focus our attention on scale-invariant models with
three-loop neutrino mass. If we also seek to address the DM problem, a
minimal approach would see the DM play a role in either generating neutrino
mass or triggering electroweak symmetry breaking. Thus, we arrive at a
picture in which \emph{both the weak scale and neutrino mass arise as
radiative effects, with the weak scale, the DM mass, and the mass scale for
the exotics that induce neutrino mass, all finding a common birth, via
dimensional transmutation}. This picture can address short-comings of the
SM, while also explaining why the exotics required in three-loop neutrino
mass models have (otherwise independent) masses of $\mathcal{O}(\mathrm{TeV}%
) $ --- a common ancestry requires that they be related to the weak scale.

In this work we present a scale-invariant model for three-loop neutrino mass
that contains a fermionic DM candidate. We explore the model in detail and
present feasible parameter space that achieves the correct DM relic
abundance, while generating viable symmetry breaking and neutrino masses ---
all compatible with low-energy constraints. As per usual for scale-invariant
frameworks, the model predicts a dilaton. However, here the dilaton has the
dual role of allowing electroweak symmetry breaking and simultaneously
sourcing the lepton number violation that allows radiative neutrino masses.
We note that a number of earlier works studied relationships between the
origin of neutrino mass and DM, see e.g.~Refs.~\cite{Krauss:2002px,
Ma:2006km, AN2013, Aoki:2013gzs, Ng:2014pqa, Culjak:2015qja}. There has also
been much interest in scale-invariant models in recent years, see e.g.~Refs.~%
\cite{Hempfling:1996ht, Iso:2009ss, Heikinheimo:2013fta, Abel:2013mya,
Kannike:2014mia, Hamada:2015ria}.

The structure of this paper is as follows. In Section~\ref%
{sec:symmetry_breaking} we introduce the model and detail the symmetry
breaking sector. We turn our attention to the origin of neutrino mass in
Section~\ref{sec:neutrino_mass} and discuss various constraints in Section~%
\ref{sec:constraints}. Dark matter is discussed in Section~\ref%
{sec:dark_matter} and our main analysis and results appear in Section~\ref%
{sec:results}. Conclusions are drawn in Section~\ref{sec:conc}.

\section{A Scale-Invariant Three-Loop Model\label{sec:symmetry_breaking}}

We consider a classically scale-invariant (SI) extension of the SM in which
neutrino mass appears at the three-loop level. The SM is extended by the
addition of two charged scalars, $S_{1,2}^+\sim(1,1,2)$, three singlet
fermions, $N_{iR}\sim(1,1,0)$, with $i\in\{1,\,2,\,3\}$ labeling
generations, and a singlet scalar, $\phi\sim(1,1,0)$.\footnote{%
Quantities in parentheses refer to quantum numbers under the SM gauge
symmetry $SU(3)_c\otimes SU(2)_L\otimes U(1)_Y$.} A $Z_2$ symmetry with
action $\{S_2,\,N_R\}\rightarrow\{-S_2,\,-N_R\}$ is imposed, with all other
fields being $Z_2$-even. This symmetry remains exact in the full theory,
making the lightest $Z_2$-odd field a stable DM candidate, which should be
taken as the lightest fermion, $N_{1}\equiv N_\text{{\tiny DM}}$, to avoid a
cosmologically-excluded stable charged particle. The scalar $\phi$ plays a
key role in triggering electroweak symmetry breaking, as explained below,
and also ensures that lepton number symmetry is explicitly broken, thereby
allowing radiative neutrino mass.

Consistent with the SI and $Z_{2}$ symmetries, the Lagrangian contains the
following terms:
\begin{equation}
\mathcal{L}\supset -\;\{f_{\alpha \beta }\,\overline{L_{\alpha }^{c}}%
\,L_{\beta }\,S_{1}^{+}+g_{i\alpha }\,\overline{N_{i}^{c}}%
\,S_{2}^{+}\,e_{\alpha R}+\mathrm{H.c}\}\;-\;\frac{1}{2}\,\tilde{y}%
_{i}\,\phi \,\overline{N_{i}^{c}}\,N_{i}\;-\;V(H,S_{1,2},\phi ),
\label{eq:SI_KNT_Lagrangian}
\end{equation}%
where Greek letters label SM flavors, $\alpha,\,\beta \in \{e,\,\mu ,\,\tau
\}$, and $f_{\alpha \beta }$, $g_{i\alpha }$ and $\tilde{y}_{i}$ are Yukawa
couplings. The $Z_{2}$ symmetry forbids the term $\bar{L}\tilde{H}N_{R}$,
which would otherwise generate tree-level neutrino masses after the SM
scalar $H\sim (1,2,1)$ develops a VEV. The potential $V(H,S_{1,2},\phi )$ is
the most-general potential consistent with the SI and $Z_{2}$ symmetries.

\subsection{Symmetry Breaking\label{sec:KNT_symmetry_breaking}}

We are interested in parameter space where both $\phi $ and $H$ acquire
nonzero vacuum expectation values (VEVs), $\langle H\rangle \neq 0$ and $%
\langle \phi \rangle \neq 0$. This breaks both the SI and electroweak
symmetries while preserving the $Z_{2}$ symmetry. The most-general scalar
potential includes the terms
\begin{equation}
V_{0}(H,\,S_{1,2},\,\phi )\supset \lambda_{\text{{\tiny H}}}\ |H|^{4}+\frac{%
\lambda_{\phi \text{{\tiny H}}}}{2}|H|^{2}\phi ^{2}+\frac{\lambda_{\phi }}{%
4}\phi ^{4}+\frac{\lambda_{\text{{\tiny S}}}}{4}%
(S_{1}^{-})^{2}(S_{2}^{+})^{2}+\sum_{a=1,2}\frac{1}{2}(\lambda_{\text{%
{\tiny H}}a}\,|H|^{2}+\lambda_{\phi a}\,\phi ^{2})|S_{a}|^{2}.
\label{eq:KNT_pot}
\end{equation}%
A complete analysis of the potential requires the inclusion of the
leading-order radiative corrections. In general the full one-loop corrected
potential is not analytically tractable. However, a useful approach for
approximating the ground state in SI models was presented in Ref.~\cite%
{Gildener:1976ih}. Taking guidance from Ref.~\cite{Gildener:1976ih}, we
adopt an approximation for the ground state that allows one to obtain simple
analytic expressions. The physical spectrum contains two charged scalars $%
S_{1,2}^{+}$, and two neutral scalars, denoted as $h_{1,2}$. As discussed in
Appendix~\ref{app:min_process}, for the present model, the minimum of the
loop-corrected potential can be approximated by neglecting loop corrections
involving only the scalars $h_{1,2}$. The viability of this simplification
follows from the dominance of the beyond-SM scalars $S_{1,2}^{+}$ (see
Appendix~\ref{app:min_process}). Adopting this approximation, the one-loop
corrected potential for the CP-even neutral scalars is
\begin{eqnarray}
V_{1-l}\left( h,\phi \right) &=&\frac{\lambda_{\text{{\tiny H}}}}{4}h^{4}+%
\frac{\lambda_{\phi \text{{\tiny H}}}}{4}\phi
^{2}h^{2}+\frac{\lambda_{\phi }}{4}\phi
^{4}+\sum_{i=all~fields}n_{i}G\left( m_{i}^{2}\left( h,\phi
\right) \right) , \label{V} \\
G\left( \eta \right) &=&\frac{\eta ^{2}}{64\pi ^{2}}\left[ \log
\frac{\eta }{\Lambda ^{2}}-\frac{3}{2}\right] ,
\end{eqnarray}%
where $\Lambda $ is the renormalization scale, $n_{i}$ are the field
multiplicities, and we employ the unitary gauge, with $H=(0,h/\sqrt{2})^{T}$%
. The sum is over all fields, neglecting the light SM fermions (all but the
top quark) and the (to be determined) neutral scalar mass-eigenstates $%
h_{1,2}$. Due to the SI symmetry, the field-dependent masses can be written
as
\begin{equation}
m_{i}^{2}\left( h,\phi \right) =\frac{\alpha_{i}}{2}h^{2}+\frac{\beta_{i}}{%
2}\phi ^{2},
\end{equation}%
the constants $\alpha_{i}$ and $\beta_{i}$ are given by%
\begin{eqnarray}
\alpha_{W} &=&\frac{g^{2}}{2},~\alpha_{Z}=\frac{g^{2}+g^{\prime 2}}{2}%
,~\alpha_{t}=y_{t}^{2},~\alpha_{S_{a}}=\lambda_{\text{{\tiny H}}%
a},~\alpha_{N_{i}}=0, \notag \\
\beta_{W} &=&\beta_{Z}=\beta_{t}=0,~\beta_{S_{a}}=\lambda_{\phi
a},~\beta_{N_{i}}=2\tilde{y}_{i}^{2},
\end{eqnarray}%
with $g$ ($g^{\prime }$) and $y_{t}$ are the $SU(2)_{L}$ ($U(1)_{Y}$) gauge
and top Yukawa couplings, respectively.

Dimensional transmutation introduces a dimensionful parameter into
the theory in exchange for one of the dimensionless couplings. In
the present model, an analysis of the potential shows that a minimum
with $\langle h\rangle \equiv v\neq 0$ and $\langle \phi \rangle
\equiv x\neq 0$ exists for $\lambda_{\phi \text{{\tiny H}}}<0$, and
is triggered at the scale where the couplings satisfy the relation
\begin{equation}
2\left\{ \lambda_{\text{{\tiny H}}}{\lambda_{\phi }}+\frac{\lambda_{\text{%
{\tiny H}}}\ }{x^{2}}\sum_{i}n_{i}\left\{ \beta_{i}-\alpha_{i}\frac{v^{2}}{%
x^{2}}\right\} G^{\prime }\left( m_{i}^{2}\right) \right\}
^{1/2}+\lambda_{\phi \text{{\tiny
H}}}+\frac{2}{x^{2}}\sum_{i}n_{i}\alpha_{i}G^{\prime }\left(
m_{i}^{2}\right) =0, \label{coupling_condition}
\end{equation}%
with $G^{\prime }\left( \eta \right) =\partial G\left( \eta \right)
/\partial \eta $. The further condition%
\begin{equation}
-\frac{\lambda_{\phi \text{{\tiny H}}}}{2\lambda_{\text{{\tiny H}}}}=\frac{%
v^{2}}{x^{2}}+\sum_{i}\frac{n_{i}\alpha_{i}}{\lambda_{\text{{\tiny
H}}}\ x^{2}}G^{\prime }\left( m_{i}^{2}\right), \label{eq:vev_cond}
\end{equation}%
is also satisfied at the minimum. Thus, for $\lambda_{\phi \text{{\tiny H}},%
\text{{\tiny H}}}=\mathcal{O}(1)$ one has $v\sim x$ and the exotic scale is
naively expected around the \textrm{TeV} scale. Note that Eqs.~%
\eqref{coupling_condition} and \eqref{eq:vev_cond} ensure that the tadpoles
vanish.\footnote{%
To our level of approximation, Eqs.~\eqref{coupling_condition} and %
\eqref{eq:vev_cond} are the loop-corrected generalizations of the
standard tree-level results, $4\sqrt{\lambda_{\text{{\tiny
H}}}(\Lambda )\,\lambda
_{\phi }(\Lambda )}+\lambda_{\phi \text{{\tiny H}}}(\Lambda )\ =\ 0$ and $%
\lambda_{\phi \text{{\tiny H}}}/2\lambda_{\text{{\tiny H}}}=v^{2}/x^{2}$~%
\cite{Foot:2007iy}.}

Defining the one-loop quartic couplings as
\begin{equation}
\lambda_{\phi }^{1-l}=\frac{1}{6}\frac{\partial
^{4}V_{1-l}}{\partial \phi ^{4}},\quad \lambda_{\text{{\tiny
H}}}^{1-l}=\frac{1}{6}\frac{\partial
^{4}V_{1-l}}{\partial h^{4}},\quad \quad \lambda_{\phi \text{{\tiny H}}%
}^{1-l}=\frac{\partial ^{4}V_{1-l}}{\partial h^{2}\partial \phi ^{2}},
\end{equation}%
vacuum stability at one-loop requires that the following conditions be
satisfied:
\begin{equation}
\lambda_{\text{{\tiny H}}}^{1-l},\lambda_{\phi }^{1-l},\lambda
_{\phi \text{{\tiny H}}}^{1-l}+2\sqrt{\lambda_{\text{{\tiny
H}}}^{1-l}\lambda_{\phi }^{1-l}}>0. \label{stability_condition}
\end{equation}%
We must also impose the condition $\lambda_{\phi \text{{\tiny
H}}}^{1-l}<0$
to ensure that the vacuum with $v\neq 0$ and $x\neq 0$ is the ground state.%
\footnote{%
For $\lambda_{\phi \text{{\tiny H}}}^{1-l}>0$ the vacuum with only
one nonzero VEV is preferred.} Eq.~\eqref{stability_condition} also
guarantees that the eigenmasses-squared for the CP-even neutral
scalars are strictly positive, and forces one of the beyond-SM
scalars $S_{1,2}^{+}$ be the heaviest particle in the spectrum.

\subsection{The Scalar Spectrum}

The mass matrix for the neutral scalars is denoted as
\begin{equation}
V_{1-l}(h,\,\phi )\supset \frac{1}{2}(h,\,\phi )\left(
\begin{array}{cc}
m_{hh}^{2} & m_{h\phi }^{2} \\
m_{h\phi }^{2} & m_{\phi \phi }^{2}%
\end{array}%
\right) \left(
\begin{array}{c}
h \\
\phi%
\end{array}%
\right),
\end{equation}%
where the mass parameters $m_{hh}$, $m_{\phi \phi }$ and $m_{h\phi }$ are
calculated from the loop-corrected potential $V_{1-l}\left( h,\phi \right) $%
. The mass eigenstates are labeled as
\begin{equation}
h_{1}\,=\,\cos \theta_{h}\,h-\sin \theta_{h}\,\phi \,,\quad
\,h_{2}=\,\sin \theta_{h}\,h+\cos \theta_{h}\,\phi \,,
\label{eq:general_scalar_eigenstates}
\end{equation}%
with the eigenvalues and mixing angles given by
\begin{eqnarray}
M_{h_{1,2}}^{2} &=&\frac{1}{2}\left\{ m_{11}^{2}+m_{22}^{2}\pm \sqrt{\left(
m_{22}^{2}-m_{11}^{2}\right) ^{2}+4m_{12}^{4}}\right\}, \notag \\
\tan 2\theta_{h} &=&\frac{2m_{12}^{2}}{m_{22}^{2}-m_{11}^{2}}.
\end{eqnarray}%
Here $h_{1}$ is a massive SM-like scalar and $h_{2}$ is a pseudo-Goldstone
boson associated with SI symmetry breaking --- the latter is massless at
tree-level but acquires mass at the loop-level. One can obtain simple
tree-level expressions for the SM-like scalar mass
\begin{equation}
M_{h_{1}}^{2}=(2\lambda_{\text{{\tiny H}}}\ -\lambda_{\phi \text{{\tiny H}}%
})v^{2}, \label{eq:higgs_mass}
\end{equation}%
and the mixing angle,
\begin{equation}
c_{h}\ \equiv \ \cos \theta_{h}\ =\
\frac{x}{\sqrt{x^{2}+v^{2}}},\quad s_{h}\ \equiv \ \sin \theta_{h}\
=\ \frac{v}{\sqrt{x^{2}+v^{2}}},\, \label{eq:thetah_phivev}
\end{equation}%
though in large regions of parameter space it is important to include loop
corrections to these expressions to obtain accurate results. In our
numerical analysis we employ the full loop-corrected expressions for the
scalar masses and mixing, as is necessary to obtain $M_{h_{2}}\neq 0$. Due
to the SI symmetry, the parameters in the model are somewhat constrained,
with $\lambda_{\phi }$ and $\lambda_{\phi \text{{\tiny H}}}$ fixed by Eqs.~%
\eqref{coupling_condition} and \eqref{eq:vev_cond} while the Higgs mass $%
M_{h_{1}}\simeq 125$~\textrm{GeV} fixes $\lambda_{\text{{\tiny
H}}}$.

The tree-level masses for the charged scalars, $S_{1,2}^{+}$, are
\begin{equation}
M_{S_{a}}^{2}=\frac{1}{2}\left\{ \lambda_{\phi a}x^{2}+\lambda_{\text{%
{\tiny H}}a}v^{2}\right\} \quad \mathrm{for}\quad a=1,2,
\end{equation}%
where $S_{1}^{+}$ and $S_{2}^{+}$ do not mix due to the $Z_{2}$ symmetry.
Note that a useful approximation for $M_{h_{2}}$ is~\cite{Gildener:1976ih}
\begin{equation}
M_{h_{2}}^{2}\simeq \tfrac{1}{8\pi ^{2}(\langle \phi \rangle ^{2}+\langle
h\rangle ^{2})}\left\{
M_{h_{1}}^{4}+6M_{W}^{4}+3M_{Z}^{4}-12M_{t}^{4}+2%
\sum_{a=1}^{2}M_{S_{a}}^{4}-2\sum_{i=1}^{3}M_{N_{i}}^{4}\right\} ,
\label{eq:pgb_mass}
\end{equation}%
which shows that one of the beyond-SM scalars $S_{1,2}$ must be the heaviest
beyond-SM state in order to ensure $M_{h_{2}}>0$.

As mentioned already, we expect the VEVs to be of a similar scale, $\langle
\phi \rangle \sim \langle h\rangle $, as evidenced by Eq.~\eqref{eq:vev_cond}%
. For completeness, however, we note that there is a technically natural
limit in which one obtains $\langle \phi \rangle \gg \langle h\rangle $.
This arises when \emph{all} the couplings to $\phi $ are taken to be
hierarchically small, namely $\{\tilde{y}_{i},\,\lambda_{\phi \text{{\tiny H%
}}},\,\lambda_{\phi 1,2}\}\ll 1$, with the masses $M_{h_{1}}$, $M_{N}$ and $%
M_{S_{1,2}}$ held at $\mathcal{O}(\mathrm{TeV})$. This feature
reflects the fact that $\phi $ decouples in the limit
$\{\tilde{y}_{i},\,\lambda_{\phi \text{{\tiny H}}},\,\lambda_{\phi
1,2}\}\rightarrow 0$, up to gravitational
effects~\cite{Foot:2013hna}. In this limit we expect the model to be
very similar to the KNT model~\cite{Krauss:2002px}, but with a
light, very weakly-coupled scalar in the spectrum, $h_{2}$. Absent a
compelling motivation for such hierarchically small parameters, we
restrict our attention to values of $\langle \phi \rangle \leq
5$~\textrm{TeV}.

\section{Neutrino Mass\label{sec:neutrino_mass}}

\begin{figure}[ht]
\begin{center}
\includegraphics[width = 0.55\textwidth]{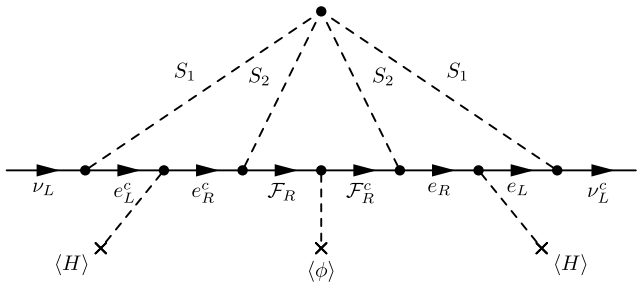}
\end{center}
\caption{Three-loop diagram for neutrino mass in a scale-invariant model.}
\label{fig:conf_3loop_nuDM}
\end{figure}

We now turn to the origin of neutrino mass. The $Z_{2}$-odd fermions, $N_{i}$%
, develop masses $M_{N_{i}}=\tilde{y}_{i}\langle \phi \rangle $, and do not
mix with SM leptons due to the $Z_{2}$ symmetry. We order their masses as $%
M_{\text{{\tiny DM}}}\equiv M_{N_{1}}<M_{N_{2}}<M_{N_{3}}$. SM neutrinos, on
the other hand, acquire mass radiatively. The combination of the Yukawa
interactions in Eq.~\eqref{eq:SI_KNT_Lagrangian} and the term
\begin{equation}
V(H,\,S_{1,2},\,\phi )\supset \frac{\lambda_{\text{{\tiny S}}}}{4}%
(S_{1}^{-})^{2}(S_{2}^{+})^{2}, \label{eq:KNT_pot_Lbreaking}
\end{equation}%
in the scalar potential, explicitly break lepton number symmetry.
Consequently neutrino masses appear at the three-loop level as shown in
Figure~\ref{fig:conf_3loop_nuDM}.

Calculating the loop diagram, the mass matrix has the form
\begin{equation}
(\mathcal{M}_{\nu })_{\alpha \beta }=\frac{\lambda_{\text{{\tiny S}}}}{%
(4\pi ^{2})^{3}}\,\frac{m_{\sigma }m_{\rho }}{M_{S_{2}}}\,g_{\sigma i}^{\ast
}\,g_{\rho i}^{\ast }\,f_{\alpha \sigma }\,f_{\beta \rho }\times
F_{loop}\left( \frac{M_{N_{i}}^{2}}{M_{S_{2}}^{2}},\frac{M_{S_{1}}^{2}}{%
M_{S_{2}}^{2}}\right),
\end{equation}%
where $m_{\sigma,\rho }$ denote charged lepton masses and the function $%
F_{loop}(x,y)$ encodes the loop integrals \cite{AN2013}
\begin{equation}
F_{loop}(\alpha,\beta )=\frac{\sqrt{\alpha }}{8\beta ^{2}}\int_{0}^{\infty
}dr\frac{r}{r+\alpha }\left( \int_{0}^{1}dx\ln \frac{x(1-x)r+(1-x)\beta +x}{%
x(1-x)r+x}\right) ^{2}.
\end{equation}

One can relate the neutrino mass matrix to the elements of the
Pontecorvo-Maki-Nakawaga-Sakata (PMNS) mixing matrix~\cite{Pontecorvo:1967fh}
elements, we parameterize the latter as
\begin{equation}
U_{\nu }=\left(
\begin{array}{ccc}
c_{12}c_{13} & c_{13}s_{12} & s_{13}e^{-i\delta_{d}} \\
-c_{23}s_{12}-c_{12}s_{13}s_{23}e^{i\delta_{d}} &
c_{12}c_{23}-s_{12}s_{13}s_{23}e^{i\delta_{d}} & c_{13}s_{23} \\
s_{12}s_{23}-c_{12}c_{23}s_{13}e^{i\delta_{d}} &
-c_{12}s_{23}-c_{23}s_{12}s_{13}e^{i\delta_{d}} & c_{13}c_{23}%
\end{array}%
\right) \times U_{m},
\end{equation}%
with $\delta_{d}$ the Dirac phase and
$U_{m}=\mathrm{diag}(1,\,e^{i\theta_{\alpha }/2},\,e^{i\theta
_{\beta }/2})$ encoding the Majorana phase dependence. The shorthand
$s_{ij}\equiv \sin \theta_{ij}$ and $c_{ij}\equiv \cos \theta_{ij}$
refers to the mixing angles. For our numerical scans (discussed
below) we fit to the best-fit experimental values for the mixing
angles and mass-squared differences: $s_{12}^{2}=0.320_{-0.017}^{+0.016}$, $%
s_{23}^{2}=0.43_{-0.03}^{+0.03}$, $s_{13}^{2}=0.025_{-0.003}^{+0.003}$, $%
|\Delta m_{13}^{2}|=2.55_{-0.09}^{+0.06}\times 10^{-3}\mathrm{eV}^{2}$ and $%
\Delta m_{21}^{2}=7.62_{-0.19}^{+0.19}\times 10^{-5}\mathrm{eV}^{2}$~\cite%
{Tortola:2012te}. Furthermore, we require that the contribution to
neutrino-less double beta decay in this model satisfies the current bound.
Within these ranges, one determines the parameter space where viable
neutrino masses and mixing occur in the model.

\section{Experimental Constraints\label{sec:constraints}}

In this section we discuss the constraints on the model from the lepton
flavor violating process $\mu \rightarrow e \gamma$, the electroweak
precision tests, the invisible Higgs decay, and the effect on $h \gamma
\gamma$ process.

\subsection{Lepton flavor}

Flavor changing processes like $\mu \rightarrow e+\gamma $ arise via loop
diagrams containing virtual charged scalars and give important constraints
on the model. At one-loop the branching ratio for $\mu \rightarrow e+\gamma $
is
\begin{eqnarray}
\mathcal{B}(\mu \rightarrow e\gamma ) &=&\frac{\Gamma (\mu \rightarrow
e+\gamma )}{\Gamma (\mu \rightarrow e+\nu +\bar{\nu})} \notag \\
&\simeq &\frac{\alpha \upsilon ^{4}}{384\pi }\times \left\{ \frac{|f_{\mu
\tau }f_{\tau e}^{\ast }|^{2}}{M_{S_{1}}^{4}}+\frac{36}{M_{S_{2}}^{4}}%
\left\vert \sum_{i}g_{ie}^{\ast }g_{i\mu
}F_{2}(M_{i}^{2}/M_{S_{2}}^{2})\right\vert ^{2}\right\},
\label{eq:muEgamma_BF}
\end{eqnarray}%
where $F_{2}(R)=[1-6R+3R^{2}+2R^{3}-6R^{2}\log R]/[6(1-R)^{4}]$. The
corresponding expression for $\mathcal{B}\tau \rightarrow \mu +\gamma )$
follows from a simple change of flavor labels in Eq.~\eqref{eq:muEgamma_BF}.
Similarly, the one-loop contributions to the anomalous magnetic moment of
the muon are%
\begin{equation}
\delta a_{\mu }=-\frac{m_{\mu }^{2}}{16\pi ^{2}}\left\{ \sum_{\alpha
\neq \mu }\frac{|f_{\mu \alpha
}|^{2}}{6M_{S_{1}}^{2}}+\sum_{i}\frac{|g_{i\mu
}|^{2}}{M_{S_{2}}^{2}}F_{2}(M_{i}^{2}/M_{S_{2}}^{2})\right\}.
\label{amu}
\end{equation}%
Null-results from searches for neutrino-less double-beta decay give an
additional constraint of $(\mathcal{M}_{\nu })_{ee}\lesssim 0.35$~eV\ \cite%
{Simkovic:2009pp}, though we find this is easily satisfied.

\subsection{Electroweak precision tests}

In principle, precision electroweak measurements can provide additional
constraints. The oblique parameters characterizing new physics effects are
given by \cite{OBL}
\begin{align}
\frac{\alpha }{4s_{W}^{2}c_{W}^{2}}\,S& =\frac{A_{ZZ}\left( M_{Z}^{2}\right)
-A_{ZZ}\left( 0\right) }{M_{Z}^{2}}-\left. \frac{\partial A_{\gamma \gamma
}\left( q^{2}\right) }{\partial q^{2}}\right\vert_{q^{2}=0}+\frac{%
c_{W}^{2}-s_{W}^{2}}{c_{W}s_{W}}\left. \frac{\partial A_{\gamma Z}\left(
q^{2}\right) }{\partial q^{2}}\right\vert_{q^{2}=0}, \label{S} \\[1mm]
\alpha T& =\frac{A_{WW}\left( 0\right)
}{M_{W}^{2}}-\frac{A_{ZZ}\left( 0\right) }{M_{Z}^{2}}. \label{T}
\end{align}%
Here, $\alpha =e^{2}/\left( 4\pi \right)
=g^{2}s_{\mathrm{w}}^{2}/\left( 4\pi \right) $ is the fine-structure
constant, $s_{\mathrm{w}}=\sin {\theta _{\mathrm{w}}}$ and
$c_{\mathrm{w}}=\cos {\theta_{\mathrm{w}}}$ are the sine and cosine,
respectively, of the Weinberg angle $\theta_{\mathrm{w}}$, and the
functions $A_{VV^{\prime }}\left( q^{2}\right) $ are the
coefficients of $g^{\mu \nu }$ in the vacuum-polarization tensors
$\Pi_{VV^{\prime }}^{\mu \nu }\left( q\right) =g^{\mu \nu
}A_{VV^{\prime
}}\left( q^{2}\right) +q^{\mu }q^{\nu }B_{VV^{\prime }}\left( q^{2}\right) $%
, where $VV^{\prime }$ could be either $\gamma \gamma $, $\gamma Z$, $ZZ$,
or $WW.$ In our model, the oblique parameters are given by \cite%
{Grimus:2008nb}
\begin{eqnarray}
\Delta T &=&\frac{3}{16\pi s_{\mathrm{w}}^{2}M_{W}^{2}}\left\{ c_{h}^{2}%
\left[ F\left( M_{Z}^{2},M_{h_{1}}^{2}\right) -F\left(
M_{W}^{2},M_{h_{1}}^{2}\right) \right] +s_{h}^{2}\left[ F\left(
M_{Z}^{2},M_{h_{2}}^{2}\right) -F\left( M_{W}^{2},M_{h_{2}}^{2}\right) %
\right] \right. \notag \\
&&\left. -\left[ F\left( M_{Z}^{2},M_{h}^{2}\right) -F\left(
M_{W}^{2},M_{h}^{2}\right) \right] \right\},
\end{eqnarray}%
\begin{eqnarray}
\Delta S &=&\frac{1}{24\pi }\left\{ 4s_{\mathrm{w}}^{4}G\left(
M_{S_{1}}^{2},M_{S_{1}}^{2},M_{Z}^{2}\right) +4s_{\mathrm{w}}^{4}G\left(
M_{S_{2}}^{2},M_{S_{2}}^{2},M_{Z}^{2}\right) +c_{h}^{2}\ln \frac{{%
M_{h_{1}}^{2}}}{{M_{h}^{2}}}+s_{h}^{2}\ln \frac{{M_{h_{2}}^{2}}}{{M_{h}^{2}}}%
\right. \notag \\
&&\left. +c_{h}^{2}\,\hat{G}\left( M_{h_{1}}^{2},M_{Z}^{2}\right)
+s_{h}^{2}\,\hat{G}\left( M_{h_{2}}^{2},M_{Z}^{2}\right) -\hat{G}\left(
M_{h}^{2},M_{Z}^{2}\right) \right\},
\end{eqnarray}%
where the functions $F$, $G$ and $\hat{G}$ are given in the appendix and $%
M_{h}=125.09$~\textrm{GeV} denotes the reference value.

\subsection{Higgs invisible decay}

The model can also face constraints from the invisible Higgs decay, $%
\mathcal{B}(h\rightarrow inv)<17\%$~\cite{Hinv}. In our case we have $%
inv\equiv \{h_{2}h_{2}\},\{N_{\text{{\tiny DM}}}N_{\text{{\tiny DM}}}\}$,
when kinematically available. The corresponding decay widths are given by%
\begin{align}
\Gamma \left( h_{1}\rightarrow h_{2}h_{2}\right) & =\frac{1}{32\pi }\frac{%
\left( \lambda_{122}\right) ^{2}}{M_{h_{1}}}\,\left( 1-\frac{4M_{h_{2}}^{2}%
}{M_{h_{1}}^{2}}\right) ^{\frac{1}{2}}\Theta \left(
M_{h_{1}}-2M_{h_{2}}\right), \notag \\
\Gamma \left( h_{1}\rightarrow N_{\text{{\tiny DM}}}N_{\text{{\tiny DM}}%
}\right) & =\frac{\tilde{y}_{\text{{\tiny DM}}}^{2}s_{h}^{2}}{16\pi }%
M_{h_{1}}\left( 1-\frac{4M_{\text{{\tiny DM}}}^{2}}{M_{h_{1}}^{2}}\right) ^{%
\frac{3}{2}}\Theta \left( M_{h_{1}}-2M_{\text{{\tiny DM}}}\right).
\end{align}%
The effective cubic coupling $\lambda_{122}$ is defined below in Eq.~%
\eqref{eq:eff_couplings}. Due to the SI symmetry, we find that $\lambda
_{122}$ vanishes at tree-level, with the small (loop-level) coupling
sufficient to ensure the decay to $h_{2}$ pairs is highly suppressed.%
\footnote{%
Note that $h_{2}$ decays to SM states, much like a light SM Higgs boson but
with suppression from the mixing angle, $s_{h}^{2}$. However, currently
there are no dedicated ATLAS or CMS searches for light scalars in the
channels $2b$, $2\tau $ or $2\gamma $, so we classify the decay $%
h_{1}\rightarrow h_{2}h_{2}$ as invisible. In practice the suppression of $%
\Gamma (h_{1}\rightarrow h_{2}h_{2})$ due to SI symmetry renders this point
moot.}

\subsection{The Higgs decay channel $h\rightarrow \gamma %
\gamma $}

The existence of extra charged scalars modifies the two Higgs branching
ratios $\mathcal{B}(h\rightarrow \gamma \gamma,\gamma Z)$, and this
deviation can be parameterized by the ratios:%
\begin{align}
R_{\gamma \gamma }& =\frac{\mathcal{B}(h\rightarrow \gamma \gamma )}{%
\mathcal{B}^{SM}(h\rightarrow \gamma \gamma )}=\left\vert 1+\frac{\upsilon }{%
2c_{h}}\frac{\frac{\vartheta_{1}}{m_{S_{1}}^{2}}A_{0}^{\gamma \gamma
}\left( \tau_{S_{1}}\right) +\frac{\vartheta_{2}}{m_{S_{2}}^{2}}%
A_{0}^{\gamma \gamma }\left( \tau_{S_{2}}\right) }{A_{1}^{\gamma
\gamma }\left( \tau_{W}\right) +N_{c}Q_{t}^{2}A_{1/2}^{\gamma \gamma
}\left( \tau
_{t}\right) }\right\vert ^{2}, \label{YY} \\
R_{\gamma Z}& =\frac{\mathcal{B}(h\rightarrow \gamma Z)}{\mathcal{B}%
^{SM}(h\rightarrow \gamma Z)}=\left\vert 1+\frac{s_{\mathrm{w}}\upsilon }{%
c_{h}}\frac{\frac{\vartheta_{1}}{m_{S_{1}}^{2}}A_{0}^{\gamma
Z}\left( \tau
_{S_{1}},\lambda_{S_{1}}\right) +\frac{\vartheta_{2}}{m_{S_{2}}^{2}}%
A_{0}^{\gamma Z}\left( \tau_{S_{2}},\lambda_{S_{2}}\right) }{c_{\mathrm{w}%
}A_{1}^{\gamma Z}\left( \tau_{W},\lambda_{W}\right) +\frac{2\left( 1-8s_{%
\mathrm{w}}^{2}/3\right) }{c_{\mathrm{w}}}A_{1/2}^{\gamma Z}\left(
\tau_{t},\lambda_{t}\right) }\right\vert ^{2}, \label{YZ}
\end{align}%
where $\tau_{X}=M_{h_{1}}^{2}/4M_{X}^{2}$ and $\lambda
_{X}=M_{Z}^{2}/4M_{X}^{2}$, with $M_{X}$ is the mass of the charged
particle $X$ running in the loop, $N_{c}=3$ is the color number,
$Q_{t}$ is the electric charge of the top quark in unit of
$\left\vert e\right\vert $, and the loop amplitudes $A_{i}$ for spin
$0$, spin $1/2$ and spin $1$ particle
contribution \cite{djouadi}, which are given in the appendix. Here $%
\vartheta_{i}$, are the SM-like Higgs couplings to the pairs of
charged
scalars $S_{1,2}^{\pm }$, which are given by%
\begin{equation}
\vartheta_{a}=c_{h}\lambda_{\text{{\tiny H}}a}v+s_{h}\lambda_{\phi
a}x.
\end{equation}

The effect of the charged scalars on (\ref{YY}) and (\ref{YZ}) depends on
the masses for $S_{a}^{\pm }$, the sign and the strength of their couplings
to the SM Higgs doublet and the neutral singlet and on the mixing angle $%
\theta_{h}$. One can use the reported results from LHC to
constraints these parameters.

\section{Dark Matter\label{sec:dark_matter}}

\subsection{Relic Density}

The lightest $Z_{2}$-odd field is a stable DM candidate. As mentioned
already, the lightest exotic fermion $N_{\text{{\tiny DM}}}\equiv N_{1}$ is
the only viable DM candidate in the model. The relic density is given by
\cite{wimps}%
\begin{equation}
\Omega_{DM}h^{2}=\frac{1.04\times 10^{9}\mathrm{GeV}^{-1}}{M_{Pl}}\frac{1}{%
\sqrt{g_{\ast }(T_{f})}<\sigma \upsilon_{r}(x_{f})>},
\label{eq:DM_Omega}
\end{equation}%
where $M_{Pl}=1.22\times 10^{19}$ $\mathrm{GeV}$\ is the Planck scale, $%
g_{\ast }(T)$ is the total effective number of relativistic particle at
temperature $T$, and%
\begin{eqnarray}
\left\langle \sigma (N_{\text{{\tiny DM}}}\ N_{\text{{\tiny DM}}%
})v_{r}\right\rangle &=&\sum_{X}\left\langle \sigma (N_{\text{{\tiny
DM}}}\ N_{\text{{\tiny DM}}}\rightarrow X)v_{r}\right\rangle
=\frac{1}{8TM_{\text{{\tiny DM}}}^{4}K_{2}^{2}\left(
\frac{M_{\text{{\tiny DM}}}}{T}\right) }
\times \notag \\
&&\int_{4M_{\text{{\tiny DM}}}^{2}}^{\infty }ds~\sigma_{N_{\text{{\tiny DM}}%
}\ N_{\text{{\tiny DM}}}\rightarrow all}(s)\left( s-4M_{\text{{\tiny DM}}%
}^{2}\right) \sqrt{s}K_{1}\left( \frac{\sqrt{s}}{T}\right) ,
\label{ThCS}
\end{eqnarray}%
is the thermally averaged DM annihilation cross-section, $v_{r}$ is the
relative velocity, $s$ is the Mandelstam variable, $K_{1,2}$ are the
modified Bessel functions and $\sigma_{N_{\text{{\tiny DM}}}\ N_{\text{%
{\tiny DM}}}\rightarrow all}(s)$ is the annihilation cross into all
kinematically accessible final state particles at the CM energy
$\sqrt{s}$.

The parameter $x_{f}=M_{\text{{\tiny DM}}}/{T_{f}}$ represents the
freeze-out temperature, and can be computed from%
\begin{equation}
x_{f}=\ln {\frac{0.03M_{Pl}M_{\text{{\tiny DM}}}<\sigma \upsilon_{r}(x_{f})>%
}{\sqrt{T_{f}}x_{f}}.} \label{eq:xf}
\end{equation}%
As will be discussed in the next section, we require that $\Omega_{N_{\text{%
{\tiny DM}}}}h^{2}$ to be in agreement with the observed value of the dark
matter relic density \cite{LCDM}.

The thermally averaged annihilation cross-section can be
approximated in the non-relativistic limit as $<\sigma
\upsilon_{r}>=a+b\upsilon_{r}^{2}$,
where $\upsilon_{r}$ is the relative DM velocity and $a$ and $b$ are the $s$%
-wave and $p$-wave factors, which receives contributions from
different annihilation channels. In this limit, the velocity squared
is approximated by $\upsilon_{r}^{2}\simeq 6/x_{f}$. Here, we
evaluate the thermally averaged cross section exactly following
(\ref{ThCS}).

\subsection{Annihilation cross section}

\begin{figure}[h]
\begin{center}
\includegraphics[width = 0.8\textwidth]{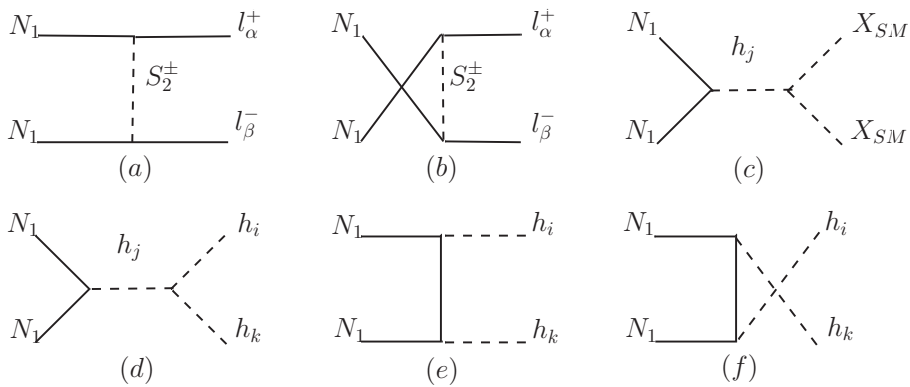}
\end{center}
\caption{Different diagrams for DM annihilation.}
\label{DM-ahn}
\end{figure}

In our model, there are many contributions, where the channels can
be classified into three types according to their Feynman diagrams
types: \textbf{(1)} annihilation into charged leptons
$N_{\text{{\tiny DM}}}N_{\text{{\tiny DM}}}\rightarrow \ell_{\alpha
}^{\mp}\ell_{\beta}^{\pm}$ (Fig. \ref{DM-ahn}-a and -b), which are
$t$-channel diagrams mediated by charged scalars\footnote{Actually,
for the same flavor case there two $s$-channel diagrams mediated by
$h_{1,2}$, however we neglect them due to the suppressed Higgs
charged leptons couplings.} \textbf{(2)} annihilation into SM
fermions and gauge
bosons pairs $N_{\text{{\tiny DM}}}N_{\text{{\tiny DM}}}\rightarrow f\bar{f}%
,~W^{-}W^{+},~ZZ$ (Fig. \ref{DM-ahn}-c), which occur through s-channel $%
h_{1,2}$-mediated diagrams, and \textbf{(3)} the annihilations into scalars,
$N_{\text{{\tiny DM}}}N_{\text{{\tiny DM}}}\rightarrow h_{1,2}h_{1,2}$ (Fig. %
\ref{DM-ahn}-d, -e and -f), which occur through both $s$- and $t$-channel
diagrams.

\textbf{Charged leptons annihilation channel}

The DM $N_{1}$ couples to SM leptons through the Yukawa couplings $%
g_{1\alpha }$, and can annihilate into charged lepton pairs as shown in Fig.~%
\ref{DM-ahn}-a\ and -b. The cross section for annihilation into charged
leptons \footnote{%
Indeed for same flavor charged leptons ($\alpha =\beta $), there are $%
h_{1,2} $ mediated $s$-channel processes that are proportional to their
Yukawa couplings; we ignore these due to the Yukawa suppression.} is given
by \cite{cheung}%
\begin{eqnarray}
\sigma (N_{\text{{\tiny DM}}}N_{\text{{\tiny DM}}} &\rightarrow
&\ell _{\alpha }^{-}\ell_{\beta }^{+})v_{r}=\frac{1}{8\pi
}\frac{|g_{1\alpha
}g_{1\beta }^{\ast }|^{2}}{s(M_{S^{+}}^{2}-M_{\text{{\tiny DM}}}^{2}+\frac{s%
}{2})^{2}}\left[ \frac{m_{\ell_{\alpha }}^{2}+m_{\ell_{\beta }}^{2}}{2}%
\left( \frac{s}{2}-M_{\text{{\tiny DM}}}^{2}\right) \right. \notag \\
&&\left[ +\frac{8}{3}\frac{(M_{S^{+}}^{2}-M_{\text{{\tiny DM}}}^{2})^{2}+%
\frac{s}{2}(M_{S^{+}}^{2}-M_{\text{{\tiny DM}}}^{2})+\frac{s^{2}}{8}}{%
(M_{S^{+}}^{2}-M_{\text{{\tiny
DM}}}^{2}+\frac{s}{2})^{2}}\frac{s}{4}\left(
\frac{s}{4}-M_{\text{{\tiny DM}}}^{2}\right) \right] , \label{Sll}
\end{eqnarray}

\textbf{SM fermions and gauge boson channels}

The processes $N_{\text{{\tiny DM}}}N_{\text{{\tiny DM}}}\rightarrow b\bar{b}
$, $t\bar{t}$, $W^{+}W^{-}$ and $ZZ$ can occur as shown in Fig. \ref{DM-ahn}%
-c. The corresponding amplitude can be written as%
\begin{equation}
\mathcal{M}=ic_{h}s_{h}y_{1}\bar{u}\left( k_{2}\right) u\left( k_{1}\right)
\left( \frac{i}{s-M_{h_{1}}^{2}}-\frac{i}{s-M_{h_{2}}^{2}}\right) \mathcal{M}%
_{h\rightarrow SM}\left( m_{h}\rightarrow \sqrt{s}\right) ,
\end{equation}%
where $\mathcal{M}_{h\rightarrow SM}\left( m_{h}\rightarrow \sqrt{s}\right) $
is the amplitude of the Higgs decay $h\rightarrow X_{SM}\bar{X}_{SM}$, with
the Higgs mass replaced as $m_{h}\rightarrow \sqrt{s}$. This leads to the
cross section%
\begin{equation}
\sigma (N_{\text{{\tiny DM}}}N_{\text{{\tiny DM}}}\rightarrow X_{SM}\bar{X}%
_{SM})\upsilon_{r}=8\sqrt{s}s_{h}^{2}c_{h}^{2}y_{1}^{2}\left\vert \frac{1}{%
s-M_{h_{1}}^{2}}-\frac{1}{s-M_{h_{2}}^{2}}\right\vert ^{2}\Gamma
_{h\rightarrow X_{SM}\bar{X}_{SM}}\left( m_{h}\rightarrow \sqrt{s}\right) ,
\label{csSM}
\end{equation}
where $\Gamma_{h\rightarrow X_{SM}\bar{X}_{SM}}\left(
m_{h}\rightarrow \sqrt{s}\right)$\ is the total decay width, with
$m_{h}\rightarrow \sqrt{s}$.

\textbf{Higgs channel}

The DM can self-annihilate to $h_{1,2}h_{1,2}$,\ as seen in Fig. \ref{DM-ahn}%
-d, -e and -f. The amplitude squared is given by%
\begin{eqnarray}
\left\vert \mathcal{M}\right\vert ^{2} &=&2\tilde{y}_{\text{{\tiny DM}}}^{2}s%
\left[ {\frac{c_{{h}}\lambda_{{1ik}}}{s-{M_{h_{1}}^{2}}}}+{\frac{s_{{h}%
}\lambda_{{2ik}}}{s-{M_{h_{2}}^{2}}}}\right] ^{2} \notag \\
&&+4c_{{i}}c_{{k}}\tilde{y}_{\text{{\tiny DM}}}^{3}M_{\text{{\tiny DM}}}%
\left[ {\frac{c_{{h}}\lambda_{{1ik}}}{s-{M_{h_{1}}^{2}}}}+{\frac{s_{{h}%
}\lambda_{{2ik}}}{s-{M_{h_{2}}^{2}}}}\right] \left( \frac{s-{M_{h_{i}}^{2}}+%
{M_{h_{k}}^{2}}}{t-M_{\text{{\tiny DM}}}^{2}}+a\frac{s+{M_{h_{i}}^{2}}-{%
M_{h_{k}}^{2}}}{u-M_{\text{{\tiny DM}}}^{2}}\right) \notag \\
&&+\frac{2c_{{i}}^{2}c_{{k}}^{2}\tilde{y}_{\text{{\tiny DM}}}^{4}}{\left(
t-M_{\text{{\tiny DM}}}^{2}\right) ^{2}}\left\{ 4M_{\text{{\tiny DM}}}^{2}{%
M_{h_{k}}^{2}}+\left( M_{\text{{\tiny DM}}}^{2}+{M_{h_{i}}^{2}}-t\right)
\left( M_{\text{{\tiny DM}}}^{2}+{M_{h_{i}}^{2}}-u\right) -s{M_{h_{i}}^{2}}%
\right\} \notag \\
&&+a^{2}\frac{2c_{{i}}^{2}c_{{k}}^{2}\tilde{y}_{\text{{\tiny DM}}}^{4}}{%
\left( u-M_{\text{{\tiny DM}}}^{2}\right) ^{2}}\left\{ 4M_{\text{{\tiny DM}}%
}^{2}{M_{h_{i}}^{2}}+\left( M_{\text{{\tiny DM}}}^{2}+{M_{h_{k}}^{2}}%
-u\right) \left( M_{\text{{\tiny DM}}}^{2}+{M_{h_{k}}^{2}}-t\right) -s{%
M_{h_{k}}^{2}}\right\} \notag \\
&&+a\frac{2c_{{i}}^{2}c_{{k}}^{2}\tilde{y}_{\text{{\tiny DM}}}^{4}}{\left(
t-M_{\text{{\tiny DM}}}^{2}\right) \left( u-M_{\text{{\tiny DM}}}^{2}\right)
}\left\{ \left( M_{\text{{\tiny DM}}}^{2}+{M_{h_{i}}^{2}}-t\right) \left( M_{%
\text{{\tiny DM}}}^{2}+{M_{h_{k}}^{2}}-t\right) \right. \notag \\
&&\left. +\left( M_{\text{{\tiny DM}}}^{2}+{M_{h_{k}}^{2}}-u\right) \left(
M_{\text{{\tiny DM}}}^{2}+{M_{h_{i}}^{2}}-u\right) -\left( s-4M_{\text{%
{\tiny DM}}}^{2}\right) \left( s-M_{h_{i}}^{2}-M_{h_{k}}^{2}\right) \right\}
,
\end{eqnarray}%
with $s$, $t$ and $u$ being the Mandelstam variables, the Yukawa couplings
defined as $\tilde{y}_{\text{{\tiny DM}}}\equiv y_{1}$, $c_{1}\equiv c_{h}$
and $c_{2}\equiv s_{h}$. Here, we integrate numerically on the phase space
in order to get the cross section for a given $s$ value. At tree-level the
effective cubic scalar couplings ($\lambda_{1ik}$ and $\lambda_{2ik})$ are
given by
\begin{eqnarray}
\lambda_{111} &=&6\lambda_{\text{{\tiny H}}}\ c_{h}^{3}v-3\lambda
_{\phi
\text{{\tiny H}}}c_{h}^{2}s_{h}v+3\lambda_{\phi \text{{\tiny H}}%
}c_{h}s_{h}^{2}x-6\lambda_{\phi }s_{h}^{3}x, \notag \\
\lambda_{112} &=&\lambda_{\phi \text{{\tiny H}}%
}c_{h}^{3}x+2c_{h}^{2}s_{h}(3\lambda_{\text{{\tiny H}}}-\lambda
_{\phi \text{{\tiny H}}})v+2c_{h}s_{h}^{2}(3\lambda_{\phi
}-\lambda_{\phi \text{{\tiny H}}})x+\lambda_{\phi \text{{\tiny H}}}s_{h}^{3}v, \notag \\
\lambda_{222} &=&\lambda_{122}\ =\ 0, \label{eq:eff_couplings}
\end{eqnarray}%
though for completeness we use the full one-loop results that can be derived
from the loop-corrected potential following \cite{AAN}. The absence of cubic
interactions $h_{1}h_{2}^{2}$ and $h_{2}^{3}$, at leading order, is a
general feature of SI models.

\subsection{Direct Detection}

Concerning direct-detection experiments, the effective low-energy Lagrangian
responsible for interactions between the DM and quarks is given by
\begin{equation}
\mathcal{L}_{N_{1}-q}^{(eff)}=a_{q}\,\bar{q}q\,N_{\text{{\tiny DM}}}^{c}N_{%
\text{{\tiny DM}}},
\end{equation}%
with%
\begin{equation}
a_{q}=-\frac{s_{h}c_{h}M_{q}M_{\text{{\tiny DM}}}}{2\upsilon x}\left[ \frac{1%
}{M_{h_{1}}^{2}}-\frac{1}{M_{h_{2}}^{2}}\right] .
\end{equation}%
Consequently, the nucleon-DM effective interaction can be written as%
\begin{equation}
\mathcal{L}_{\text{{\tiny DM}}-\mathcal{N}}^{(eff)}=a_{\mathcal{N}}\mathcal{%
\bar{N}N}N_{\text{{\tiny DM}}}^{c}N_{\text{{\tiny DM}}},
\end{equation}%
with%
\begin{equation}
a_{\mathcal{N}}=\frac{s_{h}c_{h}\left( M_{\mathcal{N}}-\frac{7}{9}M_{%
\mathcal{B}}\right) M_{\text{{\tiny DM}}}}{\upsilon x}\left[ \frac{1}{%
M_{h_{1}}^{2}}-\frac{1}{M_{h_{2}}^{2}}\right] .
\end{equation}%
In this relation, $M_{\mathcal{N}}$ is the nucleon mass and $M_{\mathcal{B}}$
the baryon mass in the chiral limit \cite{He}. Thus, the approximate
expression of the spin-independent nucleon-DM elastic cross section at low
momentum transfer reads%
\begin{equation}
\sigma_{\det }=\frac{s_{h}^{2}c_{h}^{2}M_{\mathcal{N}}^{2}\left( M_{%
\mathcal{N}}-\frac{7}{9}M_{\mathcal{B}}\right) ^{2}M_{\text{{\tiny DM}}}^{4}%
}{\pi \upsilon ^{2}x^{2}\left( M_{\text{{\tiny DM}}}+M_{\mathcal{B}}\right)
^{2}}\left[ \frac{1}{M_{h_{1}}^{2}}-\frac{1}{M_{h_{2}}^{2}}\right] ^{2}.
\end{equation}%
As will be discussed below, the most stringent constraint on
$\sigma_{det}$ comes from the present as well the recent upped bound
reported by LUX experiment \cite{Akerib:2013tjd, LUXnew}.

\section{Numerical Analysis and Results\label{sec:results}}

\begin{figure}[h]
\begin{center}
\includegraphics[width = 0.6\textwidth]{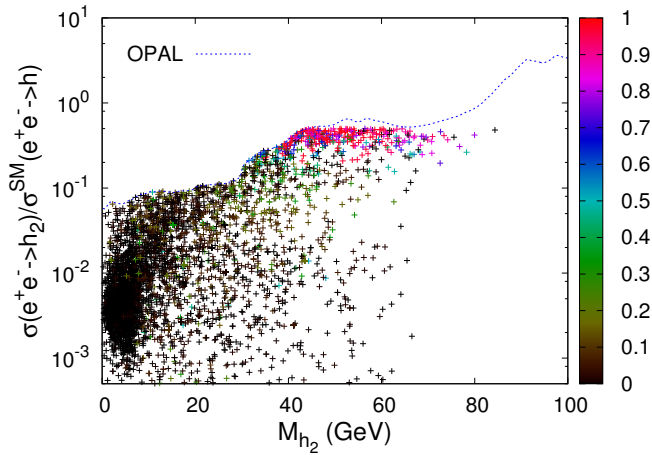}
\end{center}
\caption{Scalar mixing versus the light scalar mass. The palette gives the
branching ratio for invisible Higgs decays, with an overwhelming majority of
the points shown satisfying the constraint $B(h_{1}\rightarrow inv)<17\%$.}
\label{fig:mh2}
\end{figure}

In our numerical scan we enforce the minimization conditions, Eqs.~%
\eqref{coupling_condition} and \eqref{eq:vev_cond}, vacuum stability, the
Higgs mass $M_{h_{2}}=125.09\mp 0.21$~\textrm{GeV}, as well as the
constraints from LEP (OPAL) on a light Higgs~\cite{OPAL}. The constraint
from the Higgs invisible decay $\mathcal{B}(h\rightarrow inv)<17\%$ \cite%
{Hinv} is also enforced. All dimensionless couplings are restricted to
perturbative values and we consider the range $200~\mathrm{GeV}<\langle \phi
\rangle <5$~\textrm{TeV} for the beyond-SM VEV. We find a range of viable
values for $M_{h_{2}}$, consistent with the OPAL bounds, as shown in Figure~%
\ref{fig:mh2}. For the parameter space in our scan we tend to find $%
M_{h_{2}} $ in the range $\mathcal{O}(1)~\mathrm{GeV}\lesssim
M_{h_{2}}\lesssim90$~\textrm{GeV}. Lighter values of $M_{h_{2}} $
appear to require a degree of engineered cancelation among the
radiative mass-corrections from fermions and bosons; see
Eq.~\eqref{eq:pgb_mass}. We noticed that regions with $\langle \phi
\rangle \gtrsim 700$~\textrm{GeV} tend to be preferred in our scans.

\begin{figure}[h]
\begin{center}
\includegraphics[width = 0.5\textwidth]{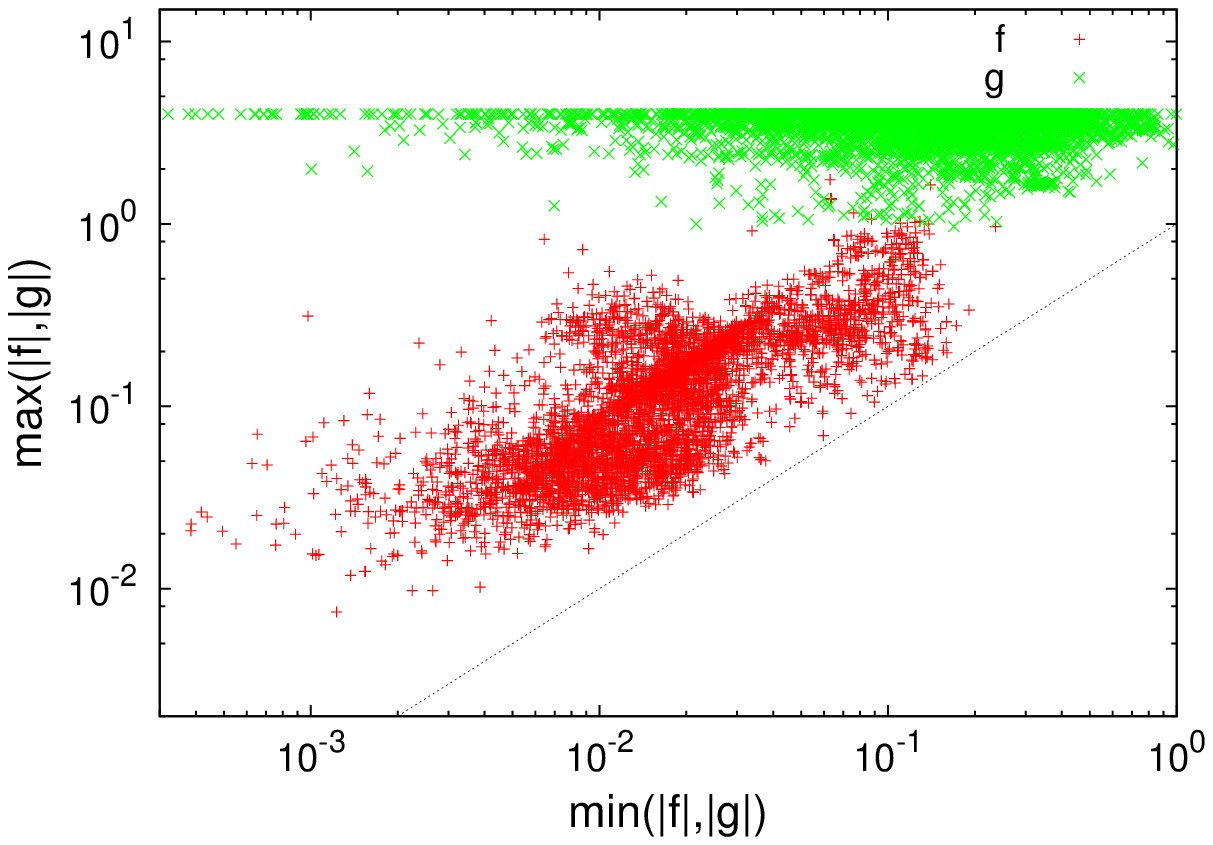}~%
\includegraphics[width =
0.5\textwidth]{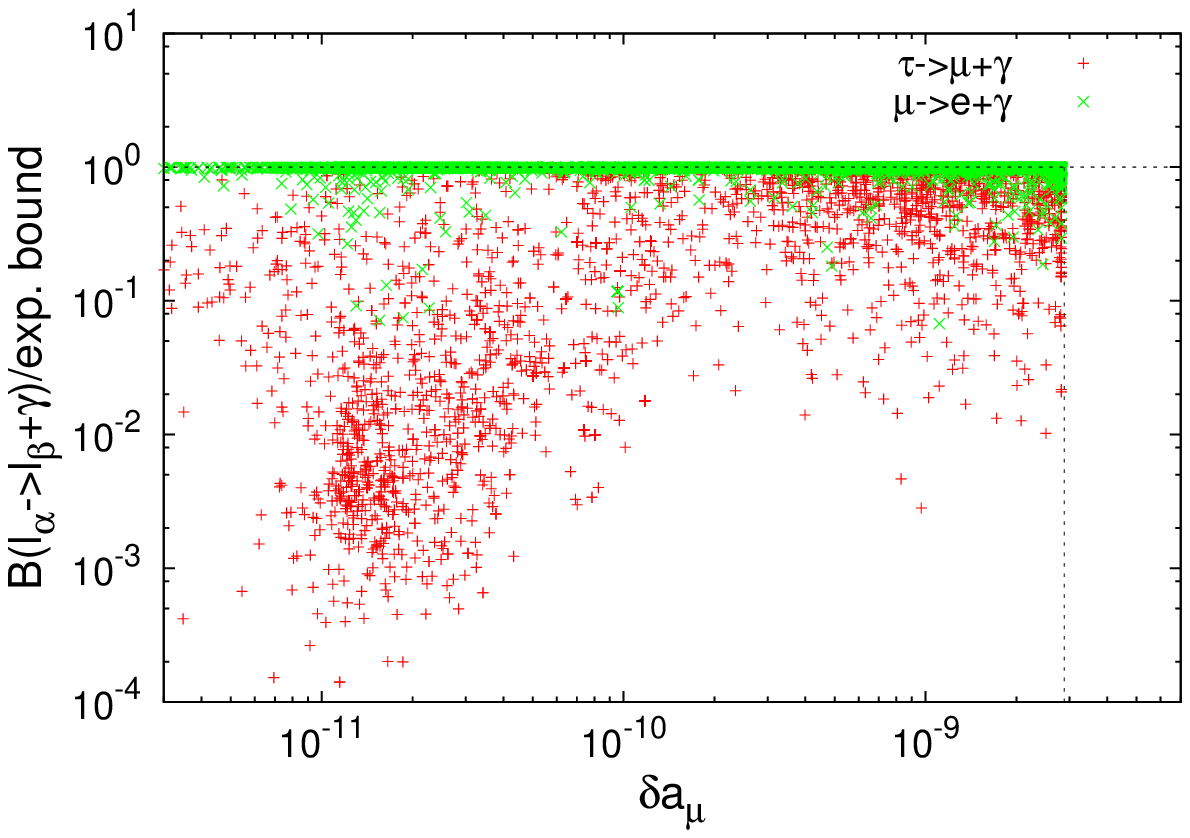}
\end{center}
\caption{Left: Viable benchmark points for the Yukawa couplings $g_{i%
\alpha }$ and $f_{\alpha \beta }$, in absolute values, where the
dashed line represents the fully degenerate case, i.e, $\min
\left\vert f\right\vert =\max \left\vert f\right\vert $. Right: The
LFV branching ratios, scaled by the experimental bounds, versus the
muon anomalous magnetic moment. The vertical line represents the
muon anomalous magnetic moment experimental constraint. }
\label{cns}
\end{figure}

We also scan for viable neutrino masses and mixing, subject to the LFV and
muon anomalous magnetic moment constraints, while also demanding a viable DM
relic density. In Figure~\ref{cns} we plot viable benchmark points for the
Yukawa couplings $g_{i\alpha}$ and $f_{\alpha\beta}$, along with the
corresponding LFV branching ratios and $\delta a_\mu$ contributions. It is
clear that the couplings $f_{\alpha\beta}$ are generally smaller than the
couplings $g_{i\alpha}$, and that the bound on $\tau\rightarrow\mu\gamma$\
is readily satisfied, while the constraint from $\mu\rightarrow e\gamma$ is
more severe. We observe in Figure~\ref{cns} that the model requires the
largest coupling in the set $g_{i\alpha}$ to take $\mathcal{O}(1)$ values.
This feature is a generic expectation for three-loop models of neutrino
mass, as one cannot make the new physics arbitrarily heavy, while reducing
the Yukawa couplings, and retain viable SM neutrino masses. Thus, the
testability of such models, which predict new physics at the TeV scale, is
generally coupled with a need for $\mathcal{O}(1)$ couplings. Consequently
one expects such couplings to encounter a Landau pole in the UV, requiring a
new description. We note that, when considering only one or two generations
of singlet fermions, no solutions that simultaneously accommodate the
neutrino mass and mixing data, low-energy flavor constraints, and the DM
relic density, were found. Therefore at least three generations of exotic
fermions are required. Also, we verified that the constraints from
neutrino-less double-beta decay searches are easily satisfied for all
benchmark points.

\begin{figure}[h]
\begin{center}
\includegraphics[width = 0.5\textwidth]{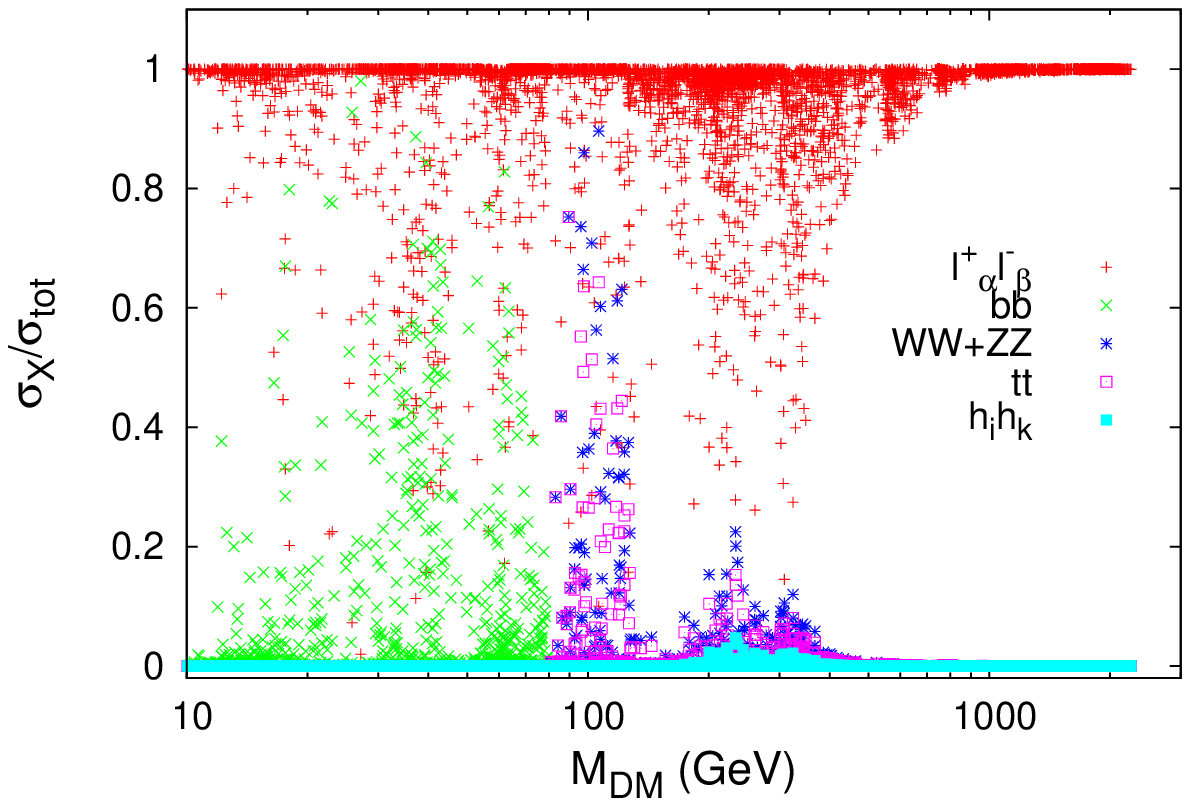}~\includegraphics[width=0.5%
\textwidth]{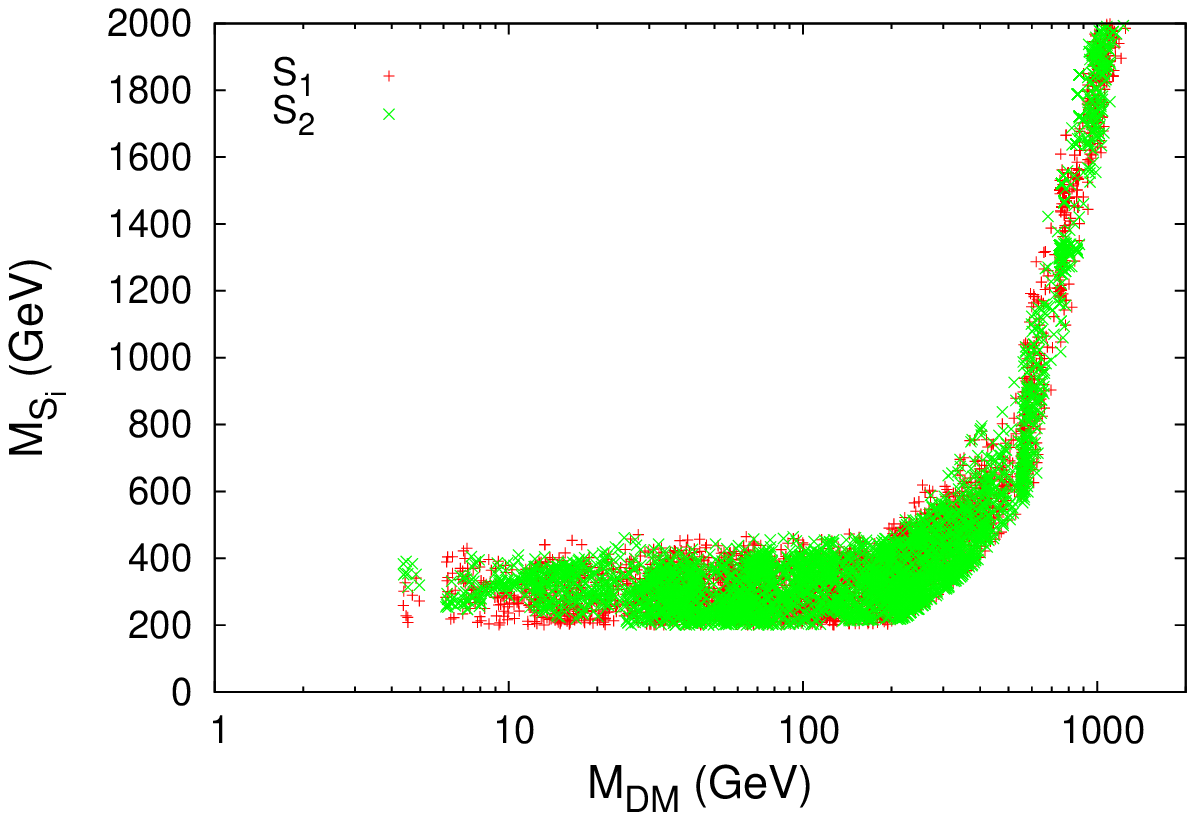}
\end{center}
\caption{Left: The relative contributions of each channel to the
annihilation cross section at the freeze-out temperature versus the
DM mass. Right: The corresponding charged scalar masses versus the
DM mass.} \label{DM}
\end{figure}

Recall that, with regards to the DM relic density, there are many classes of
annihilation channels, namely $N_{\text{{\tiny DM}}}N_{\text{{\tiny DM}}%
}\rightarrow X$ ($X=\ell_{\alpha}^{\mp}\ell_{\beta}^{\pm}$,
$b\bar{b}$, $t\bar{t}$, $WW+ZZ$, $h_{1,2}h_{1,2}$). According to the
DM mass, each channel could be significant or suppressed. In order
to probe the role of each channel, we plot the relative contribution
of each channel to the total cross section, i.e. the ratio
$\sigma_{X}/\sigma_{tot}$ at the freeze-out versus the DM
mass, in Figure~\ref{DM}-left. We see that the channel $N_{\text{{\tiny DM}}%
}N_{\text{{\tiny DM}}}\rightarrow
\ell_{\alpha}^{\mp}\ell_{\beta}^{\pm}$ is always fully dominant
except for a few benchmark points. For DM masses smaller than 80 GeV
the contribution of $X=b\bar{b}$ can be significant, while in the
range between $80$ GeV$<M_{\text{{\tiny DM}}}<100$\ GeV, both gauge bosons $%
X=WW+ZZ$ and $X=t\bar{t}$ contributions can be important. In the range $200$
GeV$<M_{\text{{\tiny DM}}}<400$\ GeV, their contribution can reach 20\%. For
large DM masses $M_{\text{{\tiny DM}}}>200$\ GeV, the $X=hh$ contribution
can reach at most 8\%. The fact that the $X=t\bar{t}$ contribution could be
important around 100 GeV, i.e., for $M_{\text{{\tiny DM}}}<M_{t}$, can be
understood due to thermal fluctuations. Figure~\ref{DM}-right shows the
corresponding charged scalar masses. For lighter DM masses of $M_{\text{%
{\tiny DM}}}<300$~\textrm{GeV}, the charged scalar masses $M_{S_{1,2}}$
should not exceed 450 \textrm{GeV}, while for larger values of $M_{\text{%
{\tiny DM}}}$, the scalar masses $M_{S_{1,2}}$ can be at the \textrm{TeV}
scale. Such light charged scalars can be within reach of collider
experiments~\cite{Ahriche:2014xra}.

\begin{figure}[h]
\begin{center}
\includegraphics[width = 0.6\textwidth]{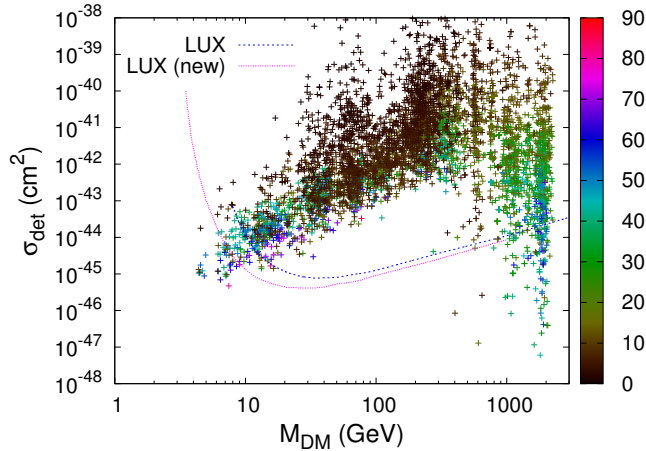}
\end{center}
\caption{The direct detection cross section versus the DM mass compared to
the recent results from LUX. The palette shows the mass for the neutral
beyond-SM scalar, $M_{h_{2}}$, in units of GeV.}
\label{det}
\end{figure}

Next we discuss the constraints from direct-detection experiments. We plot
the direct-detection cross section versus the DM mass for our benchmark
points in Figure~\ref{det}. One observes immediately that the
direct-detection limits impose serious constraints on the model, with a
large number of the benchmarks excluded by LUX~\cite{Akerib:2013tjd} as well
the improved LUX bounds \cite{LUXnew}. We find that only few benchmarks with
$M_{\text{{\tiny DM}}}\lesssim 10$~\textrm{GeV} or $M_{\text{{\tiny DM}}%
}\gtrsim 400$~\textrm{GeV} survive the LUX bounds. As is clear from the
figure, the surviving benchmarks will be subject to future tests in
forthcoming direct-detection experiments. The palette in Figure~\ref{det}
shows the corresponding values for $M_{h_{2}}$, in units of \textrm{GeV}. In
the region of parameter space for which $N_{\text{{\tiny DM}}}$ gives viable
dark matter, we find that the $M_{h_{2}}$\ must be greater than 20~GeV.

We emphasize that we only found a few benchmarks for which the DM relic
density was primarily determined by annihilations into scalars. On the
surface, this claim may appear contrary to the results of Refs.\ \cite%
{LopezHonorez:2012kv,Dutra:2015vca}, which consider Majorana DM coupled to a
singlet scalar that communicates with the SM via the Higgs portal (called
the Indirect Higgs Portal~\cite{LopezHonorez:2012kv}). Naively one may
expect our model to admit parameter space where the DM relic-density is
determined primarily by the annihilations $N_{\text{{\tiny DM}}}\ N_{\text{%
{\tiny DM}}}\rightarrow hh$, in analogy with the results of Refs.~\cite%
{LopezHonorez:2012kv,Dutra:2015vca}. However, due to the SI symmetry, our
model contains no bare mass terms, which reduces the number of free
parameters in the Lagrangian. Consequently the DM mass $M_{\text{{\tiny DM}}%
} $\ is related to both the coupling between $N_{\text{{\tiny DM}}}$ and $%
\phi $, and the mixing angle $\theta_{h}$. This reduction in
parameters means we cannot evade the LUX constraints whilst
generating a viable relic density by annihilations into scalars,
explaining the difference between our results and
Refs.~\cite{LopezHonorez:2012kv,Dutra:2015vca}. It also explains
some features of the benchmark distributions in Figure~\ref{det}.
The
benchmarks with larger contributions from the channel $N_{\text{{\tiny DM}}%
}N_{\text{{\tiny DM}}}\rightarrow hh$ have a stronger coupling between $N_{%
\text{{\tiny DM}}}$\ and $\phi $. This increases the direct-detection cross
section due to $h_{1,2}$\ exchange, creating conflict with the bounds from
LUX, so the corresponding benchmarks are strongly ruled out. Indeed, with
the smaller number of parameters in the SI model, it is a non-trivial result
that viable regions of parameter space were found in Figure~\ref{det}.

\begin{figure}[h]
\begin{center}
\includegraphics[width =0.5\textwidth]{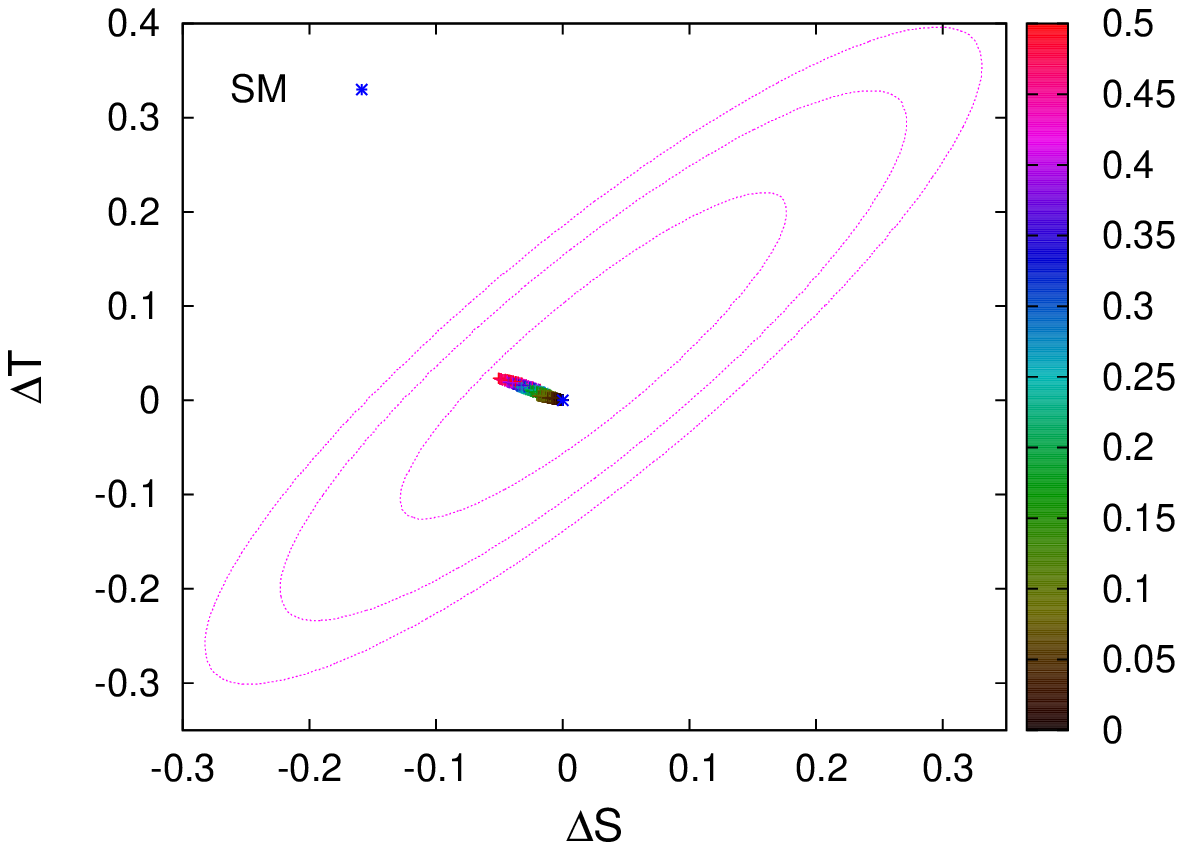}~\includegraphics[width=0.5%
\textwidth]{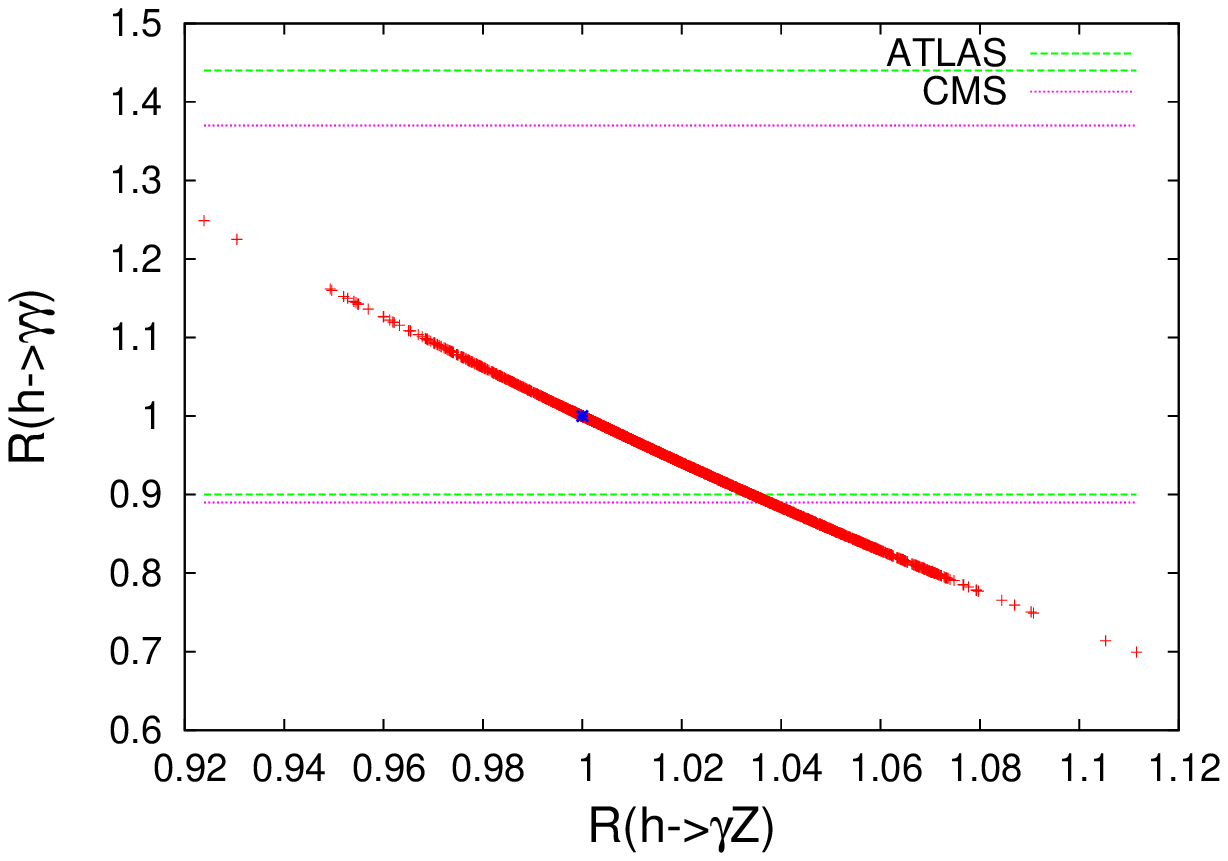}
\end{center}
\caption{Left: The oblique parameters $\Delta S$\ versus $\Delta T$\
for the benchmarks used previously. The palette shows the mixing
$\sin ^{2}\theta_{h}$ and all the points are inside the ellipsoid of
68\% CL. Right:
Ratio of the widths for $h\rightarrow \gamma \gamma $ and $%
h\rightarrow \gamma Z$ relative to the SM values. The constraints
from ATLAS and CMS are shown, along with projected sensitivities
after Run II at the LHC.} \label{hYZ}
\end{figure}

Finally, we mention that the exotics in the model allow for new
contributions to the Higgs decays $h\rightarrow \gamma \gamma $ and $%
h\rightarrow \gamma Z$. We plot the ratio of the corresponding widths
relative to the SM values in Figure~\ref{hYZ}-right. We observe that a
significant portion of the benchmarks are consistent with existing
constraints from ATLAS and CMS. Importantly, the model can be probed through
more precise measurements by ATLAS and CMS after Run II. We note that all
benchmark points are consistent with the oblique parameter constraints, as
shown in Figure~\ref{hYZ}-left.

\section{Conclusion\label{sec:conc}}

We presented a scale-invariant extension of the SM in which both the weak
scale and neutrino mass were generated radiatively. The model contains a DM
candidate, in the form of a sterile neutrino $N_{\text{{\tiny DM}}}$. A new
light neutral scalar is also predicted, namely the pseudo Goldstone-boson
associated with the broken scale-invariance, $h_{2}$, along with two charged
scalars $S_{1,2}$. The masses for the latter are generically expected to be
near the \textrm{TeV} scale, due to the related birth of the exotic scale
and the weak scale via dimensional transmutation. The constraints on the
model are rather strong, particularly the direct-detections constraints from
LUX. However, we demonstrated the existence of viable parameter space with $%
M_{\text{{\tiny DM}}}\lesssim 10$~\textrm{GeV} or $M_{\text{{\tiny DM}}%
}\gtrsim 400$~\textrm{GeV}. The model can be tested in a number of ways,
including future direct-detection experiments, collider searches for the
charged scalars, improved LFV searches, and precision measurements of the
Higgs decay width to neutral gauge bosons. We note that the model does not
possess an obvious mechanism for baryogenesis - it would be interesting to
study this matter further. In a partner paper we shall study the
scale-invariant implementation of the Ma model in Ref.~\cite{Ma:2006km}.

\section*{Acknowledgments\label{sec:ackn}}

AA wants to thank the ICTP for the hospitality during the last stage of this
work. AA is supported by the Algerian Ministry of Higher Education and
Scientific Research under the CNEPRU Project No D01720130042. KM is
supported by the Australian Research Council.

\appendix

\section{Multi-Scalar Scale-invariant Theories\label{app:min_process}}

In a general multi-scalar theory one cannot minimize the full one-loop
corrected potential analytically. However, with recourse to the underlying
SI symmetry, there exists a simple analytic approximation that captures the
leading features~\cite{Gildener:1976ih}. A general tree-level SI potential
for a set of scalars $\{\phi_A\}$ can be written as
\begin{eqnarray}
V_0(\{\phi_A\})\ =\ g_{ABCD}\, \phi_A\,\phi_B\,\phi_C\,\phi_D,
\end{eqnarray}
where the dimensionless couplings $g_{ABCD}$ are symmetric. In general,
these couplings are running parameters that depend on the energy scale, $%
g_{ABCD}=g_{ABCD}(\mu)$, and one can freely select a value of $\mu$ that
simplifies the analysis. A convenient choice is the value $\mu=\Lambda$, at
which the tree-level potential vanishes along the direction of an assumed
non-trivial minimum in field space, namely
\begin{eqnarray}
g_{ABCD}(\Lambda)\,\hat{\phi}_A\, \hat{\phi}_B\,\hat{\phi}_C\,\hat{\phi}_D=0.
\label{eq:GW_coupling_rel}
\end{eqnarray}
Here, the minimum is defined by $\langle \phi_A\rangle= R\, \hat{\phi}_A$,
with $\hat{\phi}_A$ a unit vector in field space and $R$ a (yet to be
determined) radius. Combining Eq.~\eqref{eq:GW_coupling_rel} with the
minimization conditions, $\partial V_0/\partial \phi_A=0$, determines the
angular VEVs $\hat{\phi}_A$ in terms of the couplings $g_{ABCD}$.
Subsequently expanding around the ground state in the tree-level potential
reveals a spectrum containing a massless scalar, corresponding to the flat
direction.

Eq.~\eqref{eq:GW_coupling_rel} implies that the tree-level potential
vanishes, at the scale $\mu=\Lambda$, to an accuracy on the order of the
loop corrections:
\begin{eqnarray}
V_0(\{R\hat{\phi}_A\};\mu=\Lambda)\lesssim \mathcal{O}(V_{1-\mathrm{loop}}),
\label{eq:V0_size}
\end{eqnarray}
where we display the renormalization scale dependence and write the full
loop-corrected potential as $V=V_0+V_{1-\mathrm{loop}}+\ldots$. Thus,
one-loop corrections can be comparable to $V_0$ along the direction $\hat{%
\phi}_A$, so the interplay of the two terms allows a non-trivial minimum
that lifts the flat direction to fix the radial VEV $\langle R\rangle$.
Adding the one-loop corrections along the direction $\hat{\phi}_A$ gives
\begin{eqnarray}
V(\{R\hat{\phi}_A\};\mu=\Lambda)= V_0(\{R\hat{\phi}_A\};\mu=\Lambda)+V_{1-%
\mathrm{loop}}(\{R\hat{\phi}_A\};\mu=\Lambda)+\ldots,
\label{eq:GW_pot}
\end{eqnarray}
which can be written as~\cite{Gildener:1976ih}
\begin{eqnarray}
V(\{R\hat{\phi}_A\};\mu=\Lambda)= \mathcal{A}\, R^4 +\mathcal{B}\, R^4\log%
\frac{R^2}{\Lambda^2}+\ldots, \label{eq:GW_pot2}
\end{eqnarray}
with
\begin{eqnarray}
\mathcal{A}= \frac{1}{64\pi^2\langle R\rangle^4}\left\{\mathrm{Tr}\left[%
\mathcal{M}^4_S\log\frac{\mathcal{M}^2_S}{\langle R\rangle^2}\right]-\mathrm{%
Tr}\left[\mathcal{M}^4_F\log\frac{\mathcal{M}^2_F}{\langle R\rangle^2}\right]%
+\mathrm{Tr}\left[\mathcal{M}^4_V\log\frac{\mathcal{M}^2_V}{\langle
R\rangle^2}\right]\right\}, \label{eq:A}
\end{eqnarray}
and
\begin{eqnarray}
\mathcal{B}= \frac{1}{64\pi^2\langle
R\rangle^4}\left\{\mathrm{Tr}\mathcal{M}^4_{\mathrm{S}}-\mathrm{Tr}\mathcal{M}^4_{\mathrm{F}}+
\mathrm{Tr}\mathcal{M}^4_{\mathrm{V}}\right\}.
\end{eqnarray}
Here $\mathcal{M}_{\mathrm{S},\mathrm{F},\mathrm{V}}$ are the mass matrices
for scalars, fermions and vectors, respectively, and the trace runs over
both particle species and internal degrees of freedom. Minimizing the
one-loop corrected potential lifts the flat-direction to give
\begin{eqnarray}
\langle R\rangle = e^{-\{\mathcal{A}/2\mathcal{B}+1/4\}}\Lambda.
\end{eqnarray}
The dilaton acquires a loop-level mass, given by $M^2_{\mathrm{dilaton}}=8%
\mathcal{B} \langle R\rangle^2$. Thus, radiative corrections successfully
induce a non-trivial VEV for one or more of the scalars $\phi_A$, by
introducing a dimensionful parameter, $\langle R\rangle\propto \Lambda$, in
exchange for one of the dimensionless couplings in Eq.~%
\eqref{eq:GW_coupling_rel}. This manifests dimensional transmutation.

In the present model, demanding that $M^2_{\mathrm{dilaton}}=8\mathcal{B}
\langle R\rangle^2>0$, requires that $\mathcal{B}$ be dominated by the term $%
\mathrm{Tr}\mathcal{M}^4_{\mathrm{S}}$, meaning that one (or both) of the
scalars $S^+_{1,2}$ must be the heaviest state in the spectrum. In practise,
this implies that $\mathcal{A}$ is also dominated by the contribution of $%
S^+_{1,2}$ to the $\mathcal{M}^4_{\mathrm{S}}$ term in Eq.~\eqref{eq:A}.
Thus, loop corrections from the scalars $h_{1,2}$ along the flat direction
are sub-dominant to the corrections from $S^+_{1,2}$.\footnote{%
For parameter space of interest in this work, corrections from $h_{1,2}$ are
also smaller than those from the top quark and, in large regions of
parameter space, one or more of the fermions $N$.} Therefore, simply
dropping the corrections from $h_{1,2}$ will not introduce a significant
error in the analysis (the error is expected to be $\mathcal{O}%
(M_{h_1}^4/M_{S_{1,2}}^4)$). As discussed in the text, this simplification
has the advantage of allowing one to obtain analytic expressions for the
ground state by minimizing the one-loop corrected potential directly. As a
point of comparison, for the present model, the minimization in Eq.~%
\eqref{eq:GW_coupling_rel} gives $4\sqrt{\lambda_{\text{{\tiny
H}}}(\Lambda )\,\lambda_{\phi }(\Lambda )}+\lambda_{\phi
\text{{\tiny H}}}(\Lambda ) = 0 $, and we see from
Eq.~\eqref{coupling_condition} that our approach
incorporates loop corrections to this expression, up to $\mathcal{O}%
(M_{h_1}^4/M_{S_{1,2}}^4)$ effects. Taking the heaviest scalar as
$M_S\gtrsim300$~GeV (which we can always do - see Figure~\ref{DM}),
the error in the loop terms is typically $\lesssim 3\%$. Once we
have found the ground state, we reintroduce loop corrections from
$h_{1,2}$ to determine the mass eigenvalues, reducing the error in
the expressions for the scalar masses and mixings.

\section{Oblique Parameter Functions\label{app:oblique}}

The functions employed in the calculation of the oblique parameters in
Section~\ref{sec:constraints} are defined as follows:
\begin{eqnarray}
F\left( I,J\right) &\equiv &\left\{
\begin{array}{lcl}
{\frac{I+J}{2}-\frac{IJ}{I-J}\,\ln {\frac{I}{J}}} & \Leftarrow & I\neq J, \\%
*[3mm]
0 & \Leftarrow & I=J,%
\end{array}%
\right. \\
G\left( I,J,Q\right) &\equiv &-\frac{16}{3}+\frac{5\left( I+J\right) }{Q}-%
\frac{2\left( I-J\right) ^{2}}{Q^{2}} \notag \\
&&+\frac{3}{Q}\left[ \frac{I^{2}+J^{2}}{I-J}-\frac{I^{2}-J^{2}}{Q}+\frac{%
\left( I-J\right) ^{3}}{3Q^{2}}\right] \ln {\frac{I}{J}}+\frac{r}{Q^{3}}%
\,f\left( t,r\right), \\
\hat{G}\left( I,Q\right) &=&-\frac{79}{3}+9\,\frac{I}{Q}-2\,\frac{I^{2}}{%
Q^{2}}+\left( -10+18\,\frac{I}{Q}-6\,\frac{I^{2}}{Q^{2}}+\frac{I^{3}}{Q^{3}}%
-9\,\frac{I+Q}{I-Q}\right) \ln {\frac{I}{Q}} \notag \\
&&+\left( 12-4\,\frac{I}{Q}+\frac{I^{2}}{Q^{2}}\right) \frac{f\left(
I,I^{2}-4IQ\right) }{Q},
\end{eqnarray}%
with $t\equiv I+J-Q$ and $r\equiv Q^{2}-2Q\left( I+J\right) +\left(
I-J\right) ^{2},$ and
\begin{equation}
f\left( t,r\right) \equiv \left\{
\begin{array}{lcl}
{\ \sqrt{r}\,\ln {\left\vert \frac{t-\sqrt{r}}{t+\sqrt{r}}\right\vert }} &
\Leftarrow & r>0, \\*[3mm]
0 & \Leftarrow & r=0, \\*[2mm]
{2\,\sqrt{-r}\,\arctan {\frac{\sqrt{-r}}{t}}} & \Leftarrow & r<0.%
\end{array}%
\right. \label{f}
\end{equation}

\section{Loop induced Higgs decay functions}

The functions used to evaluate the Higgs decay rate of $h\rightarrow \gamma
\gamma $ are given by%
\begin{align}
A_{0}^{\gamma \gamma }\left( x\right) & =-x^{-2}\left[ x-f\left(
x\right) \right], \notag \label{A} \\
A_{1/2}^{\gamma \gamma }\left( x\right) & =2x^{-2}\left[ x+\left( x-1\right)
f\left( x\right) \right], \notag \\
A_{1}^{\gamma \gamma }\left( x\right) & =-x^{-2}\left[ 2x^{2}+3x+3\left(
2x-1\right) f\left( x\right) \right],
\end{align}%
with%
\begin{equation}
f\left( x\right) =\left\{
\begin{array}{ccc}
\arcsin ^{2}\left( \sqrt{x}\right) & & x\leq 1 \\
-\frac{1}{4}\left[ \log \frac{1+\sqrt{1-x^{-1}}}{1-\sqrt{1-x^{-1}}}-i\pi %
\right] ^{2} & & x>1,%
\end{array}%
\right.
\end{equation}%
and those used in the decay rate of $h\rightarrow \gamma Z$ are given by%
\begin{align}
A_{0}^{\gamma Z}\left( x,y\right) & =I_{1}\left( x,y\right), \notag \\
A_{1/2}^{\gamma Z}\left( x,y\right) & =I_{1}\left( x,y\right) -I_{2}\left(
x,y\right), \notag \\
A_{1}^{\gamma Z}\left( x,y\right) & =\left[ \left( 1+2x\right) \tan
^{2}\theta_{w}-\left( 5+2x\right) \right] I_{1}\left( x,y\right)
+4\left( 3-\tan ^{2}\theta_{w}\right) I_{2}\left( x,y\right),
\end{align}%
with
\begin{equation}
I_{1}\left( x,y\right) =-\tfrac{1}{2\left( x-y\right) }+\tfrac{f\left(
x\right) -f\left( y\right) }{2\left( x-y\right) ^{2}}+\tfrac{y\left[ g\left(
x\right) -g\left( y\right) \right] }{\left( x-y\right) ^{2}},~I_{2}\left(
x,y\right) =\tfrac{f\left( x\right) -f\left( y\right) }{2\left( x-y\right) },
\end{equation}%
and
\begin{equation}
g\left( x\right) =\left\{
\begin{array}{ccc}
\sqrt{x^{-1}-1}\arcsin \left( \sqrt{x}\right) & & x\leq 1 \\
\frac{\sqrt{1-x^{-1}}}{2}\left[ \log \frac{1+\sqrt{1-x^{-1}}}{1-\sqrt{%
1-x^{-1}}}-i\pi \right] & & x>1.%
\end{array}%
\right.
\end{equation}

\end{document}